# Accurate Measurements of Cross-plane Thermal Conductivity of Thin Films by Dual-Frequency Time-Domain Thermoreflectance (TDTR)

Puqing Jiang, Bin Huang, and Yee Kan Koh[a]

*Department of Mechanical Engineering, National University of Singapore, Singapore 117576*

Accurate measurements of the cross-plane thermal conductivity $\Lambda_{\text{cross}}$ of a high-thermal-conductivity thin film on a low-thermal-conductivity ($\Lambda_s$) substrate (e.g., $\Lambda_{\text{cross}}/\Lambda_s > 20$) are challenging, due to the low thermal resistance of the thin film compared to that of the substrate. In principle, $\Lambda_{\text{cross}}$ could be measured by time-domain thermoreflectance (TDTR), using a high modulation frequency $f_h$ and a large laser spot size. However, with one TDTR measurement at $f_h$, the uncertainty of the TDTR measurement is usually high due to low sensitivity of TDTR signals to $\Lambda_{\text{cross}}$ and high sensitivity to the thickness $h_{\text{Al}}$ of Al transducer deposited on the sample for TDTR measurements. We observe that in most TDTR measurements, the sensitivity to $h_{\text{Al}}$ only depends weakly on the modulation frequency $f$. Thus, we performed an additional TDTR measurement at a low modulation frequency $f_0$, such that the sensitivity to $h_{\text{Al}}$ is comparable but the sensitivity to $\Lambda_{\text{cross}}$ is near zero. We then analyze the ratio of the TDTR signals at $f_h$ to that at $f_0$, and thus significantly improve the accuracy of our $\Lambda_{\text{cross}}$ measurements. As a demonstration of the dual-frequency approach, we measured the cross-plane thermal conductivity of a 400-nm-thick nickel-iron alloy film and a 3-µm-thick Cu film, both with an accuracy of ~10%. The dual-frequency TDTR approach is useful for future studies of thin films.

---

[a] Author to whom correspondence should be addressed. Electronic mail: mpekyk@nus.edu.sg

# I. INTRODUCTION

Planar structures such as thin films are commonly found in modern devices for existing and emerging electronic,[1] optoelectronic,[2] thermal insulating,[3] and thermoelectric[4] applications. In these applications, knowledge of the cross-plane thermal conductivity is crucial for designing more efficient materials[4] or improving the thermal management of the devices.[5] In addition to the technological importance, knowledge of cross-plane thermal conductivity of thin films and superlattices is also crucial for studying heat conduction at nanoscale.[6] For example, measurements of cross-plane thermal conductivity of superlattices advance our knowledge of heat transfer by coherent and incoherent phonons across superlattices.[7] Moreover, measurements of cross-plane thermal conductivity is particularly important to understand heat transport in novel materials (e.g., group III-nitrides) that cannot be grown into a high quality thick film.[8]

Two most popular techniques to measure the cross-plane thermal conductivity ($\Lambda_{cross}$) of thin films are the differential 3ω method[9,10] and the time-domain thermoreflectance (TDTR),[11,12] see for example Ref. [13] for a comparison of both techniques. In both techniques, samples are heated periodically at the surface, either electrically by a metal line (the differential 3ω method) or optically by a laser beam (TDTR). The periodic temperature oscillations at the surface of the samples induced by the heating are then monitored via either the change of electrical resistance of the same metal line (the differential 3ω method) or the change of the intensity of a reflected probe beam (TDTR). Due to the periodic heating, measurements using both approaches are only sensitive to the material properties of the samples within a distance from the surface in which the amplitude of temperature oscillation is substantial, usually called the thermal penetration depth $d_p$. For thin films, $d_p = \sqrt{D_f/\pi f}$, where $f$ is the frequency of the periodic heating at the surface, and $D_f$, $\Lambda_f$ and $C_f$ are the thermal diffusivity, thermal conductivity and volumetric heat capacity of the thin films respectively; $D_f = \Lambda_f/C_f$. The cross-plane thermal conductivity of the thin films is then derived by comparing the temperature responses to calculations of a diffusive thermal model.

There are, however, significant differences between the differential 3ω method and TDTR. One notable difference is the frequency range in which the periodic heating is applied. TDTR typically operates in the frequency range of $0.1 \leq f \leq 20$ MHz, while the differential 3ω method works at much lower modulation frequencies of $0.1 \leq f \leq 10$ kHz. Due to the low frequency applied in the 3ω method, temperature oscillations measured using the 3ω method are always sensitive to the thermal properties of substrates. Thus, for $\Lambda_{cross}$ measurements of thin films, a differential approach[9] is usually applied to isolate out the temperature response due to the thin films from that due to the substrates. As a result, the capability of the differential 3ω method to measure the thermal conductivity of thin films is quite limited, especially if the thermal conductivity of thin films is higher than that of substrates. For an insulating thin films, the minimum film thickness measurably by the 3ω method can be derived[13] as $h_{min}^{3\omega} = b\Lambda_{cross}/\Lambda_s$, where $b$ is the half width of the metal line and $\Lambda_s$ is the thermal conductivity of the substrate. Typically, $b \approx 10$ μm. Thus, even for the case of $\Lambda_{cross} = \Lambda_s$, the differential 3ω method can only be applied to measure films with thickness >10 μm.

On the other hand, for TDTR, heating at the surface of the samples is modulated at a radio frequency of up to ≈20 MHz, the maximum modulation frequency achievable for a typical TDTR setup. Measurements at $f$>20 MHz are challenging due to weaker out-of-phase signals and higher noise. Usually, to measure $\Lambda_{\text{cross}}$ of thin films with high accuracy, we choose a modulation frequency such that the TDTR measurements are sensitive to the thermal conductivity of the thin films, but not the substrate. For high-thermal-conductivity thin film on a low-thermal-conductivity substrate, this translates to choosing $f$ so that $d_p=h_f/2$, see Figs. 2 and 3 below for the explanation of this choice. From the figures and the discussions below, we find that the minimum film thickness that can be accurately measured by TDTR is roughly $h_{\min}^{TDTR} = 1.5 d_p$. For a crystalline film with $D=10^{-4}$ m$^2$ s$^{-1}$, there are $d_p=1.3$ μm and $h_{\min}^{TDTR}=2$ μm, at $f=20$ MHz.

In this paper, we develop a dual-frequency TDTR approach to extend the capability of TDTR to measure the thermal conductivity of thermally thin films. By performing an additional TDTR measurement at a lower modulation frequency $f_0$, the new approach could be used to measure $\Lambda_{\text{cross}}$ of films with thickness up to $\approx 0.85 d_p$, $\approx 1.8$ times thinner than the limit of the conventional TDTR. We demonstrate the capability of our dual-frequency approach by measuring $\Lambda_{\text{cross}}$ of a 400-nm-thick nickel-iron alloy film and a 3-μm-thick Cu film, both deposited on thermal SiO$_2$. Our dual-frequency TDTR measurements compare favorably with the thermal conductivity estimated from independent electrical resistivity measurements using a four-point probe. We discuss a guideline on the implementation of the dual-frequency approach.

## II. EXPERIMENTAL METHODS

### A. Time-domain thermoreflectance (TDTR)

Our time-domain thermoreflectance (TDTR) setup is similar to the TDTR setups in other laboratories.[11,14] A schematic diagram of our setup is shown in Fig. 1. In our TDTR setup, a Ti:sapphire laser oscillator produces a train of 150 fs laser pulses at a repetition rate of 80 MHz. The ultrashort laser pulses are split into a pump and a probe beams, cross-polarized to each other by a polarizing beam splitter (PBS). We modulate the pump beam by a radio-frequency (rf) electro-optic modulator (EOM) at a modulation frequency $f$, usually in the range of 100 kHz to 20 MHz. We modulate the probe path by an audio-frequency (af) mechanical chopper at 200 Hz to facilitate the removal of background signals due to coherent pick-ups. We adjust the delay time $t_d$ between pump and probe pulses by changing the optical path of the pump beam using a 60 cm long mechanical stage along the pump path, see Fig. 1. The delay of the pump beam introduces a phase shift of $\exp(i2\pi f t_d)$, where $i = \sqrt{-1}$. We use a single long-working-distance objective lens to focus both the pump and probe beams on the sample surface. We measure the root-mean-square (rms) average of the $1/e^2$ radii of pump and probe beams by spatial autocorrelation; details of this method are described in Ref [15]. We use different objective lenses to achieve different laser spot sizes on the samples; 20x, 10x, 5x and 2x objective lenses correspond to $1/e^2$ radii of 3 μm, 6 μm, 12 μm and 30 μm, respectively.

To prepare the samples for TDTR measurements, we deposit a layer of 100 nm thick Al film on our samples (e.g., thin films) as a transducer. During the measurements, the modulated pump beam is absorbed by the transducer layer, and periodically heats the sample at a modulation frequency $f$. The periodic temperature response at the surface of the sample is then monitored via changes of the intensity of the reflected probe beam measured by a photodiode detector. We reduce the strong signal at the laser repetition rate of 80 MHz using a 30 MHz low-pass filter and eliminate the signals at higher harmonics of $f$ using an inductor-capacitor (LC) resonant circuit. The signal at the modulation frequency $f$ is then picked up by an rf lock-in amplifier. We usually extract the thermal conductivity of the sample and the thermal conductance of the Al/sample interface from TDTR measurements by comparing the ratio of in-phase $V_{in}$ and out-of-phase $V_{out}$ signals of the lock-in amplifier at $f$, $R_f = -V_{in}/V_{out}$, to calculations of a diffusive thermal model.[16]

We routinely perform sensitivity analysis to estimate the uncertainty of our TDTR measurements. The sensitivity of TDTR signal $R_f$ to an input parameter $\alpha$ is defined as[13]

$$S_\alpha = \frac{\partial \ln R_f}{\partial \ln \alpha} \tag{1}$$

The accuracy of TDTR measurements depend on the sensitivity and accuracy of the input parameters $\alpha$ of the thermal model, including the laser spot size $w_0$, the thermal conductance $G$ of interfaces, and the thickness $h$, volumetric heat capacity $C$, cross-plane and the in-plane thermal conductivity of each layer of the sample. In addition, TDTR measurements are also affected by uncertainty in determining the phase in the reference channel of the rf lock-in amplifier $\delta\phi$. In TDTR measurements, we determine the right phase by adjusting the absolute value of the phase in the reference channel of the rf lock-in amplifier such that $V_{out}$ is constant across zero delay time. The accuracy of this procedure is estimated from the rms noise of $V_{out}$ (i.e., $\delta V_{out}$) in the short delay time range divided by the $V_{in}$ jump at 0 ps (i.e., $\Delta V_{in}$), $\delta\phi = \delta V_{out}/\Delta V_{in}$.[13] We follow Ref. [13] to set $S_\phi = R_f + 1/R_f$. Assuming that all the aforementioned uncertainties are random and independent, the uncertainty of $\Lambda$ of the sample derived from TDTR measurements is thus estimated as

$$\frac{\delta\Lambda}{\Lambda} = \sqrt{\sum_\alpha \left(\frac{S_\alpha}{S_\Lambda}\frac{\delta\alpha}{\alpha}\right)^2 + \left(\frac{S_\phi}{S_\Lambda}\delta\phi\right)^2} \tag{2}$$

Among all sources of uncertainty, thickness of the Al film $h_{Al}$ usually contributes the most to the uncertainty of the thermal conductivity derived from TDTR because of the high sensitivity of $R_f$ to $h_{Al}$. The uncertainty due to $h_{Al}$ would dominate even more significantly when measuring the cross-plane thermal conductivity $\Lambda_{cross}$ of high-thermal-conductivity thin films on low-thermal-conductivity substrates. We need a new approach that could reduce the sensitivity to $h_{Al}$ and thus improve the uncertainty of $\Lambda_{cross}$ measurements.

**B. Dual-frequency TDTR**

In this paper, we develop a dual-frequency TDTR approach to improve the accuracy of $\Lambda_{cross}$ measurements of high-thermal-conductivity thin films on low-thermal-conductivity substrates. To achieve this goal, we carefully evaluate the sensitivity of TDTR signals of a hypothetical sample of a 500 nm thick film with thermal diffusivity of $D_f=10^{-5}$ m$^2$ s$^{-1}$ on a SiO$_2$ substrate. We choose the sample geometry to match the NiFe metal film that we use to validate the dual-frequency approach, see the discussion in Section II (C) for the rationale for the choice of the validation sample. In the calculations, we fix the $1/e^2$ radii of the laser beams at $w_0=28$ µm so that the heat transfer is primarily one-dimensional and thus the TDTR signals are not sensitive to the in-plane thermal conductivity.

We plot the sensitivity of TDTR signals of the hypothetical sample as a function of modulation frequency $f$ in Fig. 2(a), with the delay time fixed at 100 ps. We find that due to high thermal conductivity of the thin film, TDTR signals are always more sensitive to $h_{Al}$ than to $\Lambda_{cross}$. Within the range of 6<$f$<20 MHz, the sensitivity to $\Lambda_{cross}$ decreases drastically as $f$ decreases, see Fig. 2(a). This is because as $f$ decreases, TDTR probes much deeper into the hypothetical sample, and since the thermal conductivity of the substrate is much lower than that of the thin film, the TDTR signals are predominantly determined by the thermal properties of the substrate. Thus, conventionally, to measure $\Lambda_{\text{cross}}$ of the thin film, TDTR is performed at the highest modulation frequency $f_h$ achievable by the TDTR setup. Often, due to lower signals and higher noise at high frequencies, $f_h$ is limited to ≈20 MHz. Thus, the limit of thinnest films measurable by the conventional TDTR is then about 1.5 times the thermal penetration depth at $f_h$, see discussion below how we derive the factor 1.5.

On the contrary, we observe that within the same modulation frequency range of 6<$f$<20 MHz, the sensitivity of TDTR signals to $h_{\text{Al}}$ depends only weakly on the modulation frequency $f$, see Fig. 2(a). We thus take advantage of this observation and propose that the accuracy of $\Lambda_{\text{cross}}$ measurements can be significantly improved by an additional TDTR measurement at a lower frequency, e.g., $f_0$≈6 MHz for the hypothetical sample. In this dual-frequency approach, we calculate $R_{dual} = R_{f_h}/R_{f_1}$ at each delay time from TDTR measurements at individual frequencies of $f_h$ and $f_0$. We then derive $\Lambda_{\text{cross}}$ of the thin film by comparing the derived $R_{dual}$ to calculations of the same thermal model for the conventional TDTR. Similar to the analysis of TDTR signals, we treat $\Lambda_{\text{cross}}$ of the thin film and the thermal conductance of Al/thin film interface as the only two free parameters. We note that dual-frequency and frequency-dependent TDTR approaches have been previously employed for measurements on thin films and bulk materials.[15,17] However, in those instances, measurements at low frequency were used to derive the heat capacity of the films, not to improve the thermal conductivity of $\Lambda_{\text{cross}}$ measurements as proposed here.

We demonstrate the advantages of the dual-frequency approach by plotting the sensitivity of $R_{dual}$ in Fig. 2(b). The sensitivity of $R_{dual}$ is defined as

$$S_\alpha^{dual} = \frac{\partial \ln(R_{f_h}/R_{f_0})}{\partial \ln \alpha} = \frac{\partial \ln R_{f_h}}{\partial \ln \alpha} - \frac{\partial \ln R_{f_0}}{\partial \ln \alpha} = S_\alpha^{\text{TDTR}}\bigg|_{f_h} - S_\alpha^{\text{TDTR}}\bigg|_{f_0} \qquad (3)$$

We find that by analyzing $R_{dual}$, the sensitivity to $\Lambda_{\text{cross}}$ is maintained, while the sensitivity to $h_{\text{Al}}$ is greatly reduced, see Fig. 2 (b). Thus, the $\Lambda_{\text{cross}}$ of the thin film can be accurately determined even though sensitivity to $\Lambda_{\text{cross}}$ is still low, because the largest source of uncertainty ($h_{\text{Al}}$) is essentially eliminated. We note that $R_{dual}$ is also moderately sensitivity to $h_{\text{f}}$. This, however, does not significantly affect the accuracy of the derived $\Lambda_{\text{cross}}$ because $h_{\text{f}}$ can be accurately determined by in-situ picosecond acoustics during the measurements.

To develop a general guideline to facilitate the choice of $f_{\text{h}}$ and $f_0$ in future experiments using the dual-frequency approach, we systematically modify the thermal diffusivity $D_{\text{f}}$ and thickness $h_{\text{f}}$ of the hypothetical thin film, and calculate the sensitivity of TDTR measurements using a wide range of modulation frequencies $f$, see Fig. 3 (a). We plot the calculated results as ratios of the sensitivity to $\Lambda_{\text{cross}}$ and that to $h_{\text{Al}}$, $S_{\Lambda_{\text{cross}}}/S_{h_{\text{Al}}}$, which roughly correspond to the accuracy of the conventional TDTR measurements, and ratios of the sensitivity to $\Lambda_{\text{s}}$ and that to $h_{\text{Al}}$, $S_{\Lambda_{\text{s}}}/S_{h_{\text{Al}}}$. We find that when we plot $S_{\Lambda_{\text{cross}}}/S_{h_{\text{Al}}}$ as a function of $h_{\text{f}}/d_p$, calculations over a wide range of film thickness, thermal properties and modulation frequencies agree quite well, see Fig. 3 (a). We find that at $d_p = 0.5 h_{\text{f}}$, TDTR measurements have the highest sensitivity to $\Lambda_{\text{cross}}$. We thus recommend that the high frequency $f_{\text{h}}$ is set either such that $d_p = 0.5 h_{\text{f}}$ or at the highest modulation frequency that could be achieved using the setup (typically ≈20 MHz), whichever is smaller. This is also the frequency to be used for measurements of thin films using the conventional TDTR approach. For the low frequency $f_0$, we recommend a modulation frequency that gives $d_p \approx 1.5 h_{\text{f}}$ because at this frequency, TDTR measurements have near zero sensitivity to both $\Lambda_{\text{cross}}$ and $\Lambda_{\text{s}}$ but still high sensitivity to $h_{\text{Al}}$, see Fig. 3 (a). We note that if we use a lower $f_0$, e.g., one that gives $d_p \approx 3 h_{\text{f}}$, TDTR measurements will be highly sensitive to $\Lambda_{\text{s}}$ and as a result, the uncertainty of the dual-frequency approach will be high. Thus, using these two recommended frequencies, $R_{dual}$ is sensitive to $\Lambda_{cross}$ but not $h_{Al}$ and $\Lambda_s$.

The dual-frequency is particularly useful to improve the accuracy of $\Lambda_{cross}$ measurements of thin films with thickness $0.85d_p<h_f<1.5d_p$, where $d_p$ is the thermal penetration depth calculated using the thermal diffusivity of the thin film at the modulation frequency $f_h$. To illustrate this point, we estimate the uncertainty of the derived $\Lambda_{cross}$ of the hypothetical thin film by the dual-frequency approach, following the general guideline that we have proposed in choosing $f_h$ and $f_0$, and compare with the uncertainty by the conventional TDTR approach in Fig. 3 (b), over a wide range of film thickness and thermal diffusivity of the film. We note that when the film is sufficiently thick ($h_f>1.5d_p$), the thin film can be measured by the conventional TDTR with reasonable accuracy due to acceptably high $S_{\Lambda_{cross}}/S_{h_{Al}}$. Within this range, the dual-frequency TDTR could be applied to improve the accuracy of the measurements, but the improvement is moderate. On the other hand, when the film is too thin (roughly $h_f<0.85d_p$), TDTR does not have enough sensitivity to measure $\Lambda_{cross}$. In this case, accurate measurements could not be achieved even with the dual-frequency approach. For films with thickness $0.85d_p<h_f<1.5d_p$, the improvement using the dual-frequency approach is drastic. For example, when $h_f=0.85d_p$ and $D_f=10^{-5}$ m$^2$ s$^{-1}$, the dual-frequency approach improves the accuracy of cross-plane thermal conductivity measurements from 32% to 15%, see Fig. 3(b).

We further test the limits of the dual-frequency TDTR approach by calculating the uncertainties of $\Lambda_{cross}$ measurements on two hypothetical thin films on 100 nm SiO$_2$ on Si. For the first film, we set $D_f=10^{-4}$ m$^2$ s$^{-1}$ and $h_f=1260$ nm, while for the second film, we set $D_f=10^{-5}$ m$^2$ s$^{-1}$ and $h_f=400$ nm; in both films, $d_p=h_f$ at $f_h=20$ MHz, the highest frequency achievable in a typical TDTR setup. We fix $f_0$ such that $d_p \approx 1.5 h_f$, according to the proposed general guideline for $f_0$. We plot the derived uncertainties as a function of a wide range of parameters in Fig. 4. For these $D_f$ and $h_f$, we find that the uncertainty of $\Lambda_{cross}$ derived using our dual-frequency approach is insensitive to the thermal conductivity of the transducer film and the thermal conductance of film/substrate interface $G_2$, see Figs. 4(a) and 4(c) respectively. For the thermal conductance of the transducer/film interface, $G_1$, we find that the proposed dual-frequency approach only improves the accuracy of $\Lambda_{cross}$ measurements when the thermal resistance of the interface is smaller than the thermal resistance of the film, $G_1 > \Lambda_{cross}/h_f$, see Fig. 4(b). When $G_1$ is small, heat dissipation from the transducer thin film is mainly impeded by the interface, and thus sensitivity to $\Lambda_{cross}$ is significantly reduced. Finally, for the thermal conductivity of the substrate $\Lambda_s$, our dual-frequency approach is only useful if $\Lambda_{cross}/\Lambda_s > 10$, see Fig. 4(d), as originally developed for.

**C. SAMPLE PREPARATION**

We test the validity of the dual-frequency TDTR approach using a 400-nm-thick $Ni_{80}Fe_{20}$ alloy film and a 3-μm-thick Cu film deposited on thermal $SiO_2$ substrates. We choose the NiFe alloy film and the Cu film for the validation because we can independently verify the thermal conductivity of the metal films from the in-plane electrical resistivity using the Wiedemann-Franz law. This comparison is justified even though the thermal and electrical measurements are not in the same direction, because the thermal conductivity of the metal film is isotropic and not affected by scattering of interfaces due to the short mean-free-paths of the heat carrier (i.e., electrons) on the order of 10 nm. Moreover, due to the short mean-free-paths of heat carriers, TDTR measurements on the metal films are not affected by the frequency dependence artifacts observed in dielectrics and semiconductors.[18]

We deposited the NiFe film by thermal evaporation with a base pressure of $10^{-8}$ Torr. We confirm the composition of the $Ni_{80}Fe_{20}$ alloy film by particle-induced X-ray emission (PIXE) measurements to an uncertainty of <0.8%. Based on the virtual crystal approximation, the volumetric heat capacity of $Ni_{80}Fe_{20}$ is estimated as $0.8C_{Ni}+0.2C_{Fe}=3.88\times10^6$ J m$^{-3}$ K$^{-1}$. We determine the thickness $h_f$ of the alloy film as 403±20 nm by Rutherford Backscattering Spectrometry (RBS). This thickness is verified by picoseconds acoustic using a sound velocity of 5500 m s$^{-1}$.[19]

We deposited the Cu film by magnetron sputtering. To determine the film thickness $h_f$, we fabricated a sharp step for the metal film by photolithography, and measured the thickness as 3044±150 nm by Atomic Force Microscopy (AFM). This thickness combined with picoseconds acoustic time interval yields a sound velocity of 4330 m s$^{-1}$ for our sputtered Cu film.

## III. RESULTS AND DISCUSSIONS

(a) $Ni_{80}Fe_{20}$ alloy film

Based on the general guidelines outlined in Section II (B), we choose $f_h$=17.4 MHz and $f_0$=4 MHz for the dual-frequency TDTR measurement of our $Ni_{80}Fe_{20}$ film. We used $f_h$=17.4 MHz because this is the maximum frequency that can be achieved with a decent signal-to-noise ratio using our TDTR setup.

We find that our measurements are also slightly sensitive to the thermal conductance $G_1$ of the Al/$Ni_{80}Fe_{20}$ interface and $G_2$ of the $Ni_{80}Fe_{20}$/thermal SiO$_2$ interface, see Fig. 2 (a). To independently measure $G_2$, we deposited a 70-nm-thick $Ni_{80}Fe_{20}$ film on a thermal SiO$_2$ substrate under the same evaporation conditions. With the $Ni_{80}Fe_{20}$ film acting as a transducer, we measured that $G_2$=27 MW m$^{-2}$ K$^{-1}$ using TDTR, with an uncertainty of ±22%.

We fit $R_{dual}$ with $\Lambda_{cross}$ of the $Ni_{80}Fe_{20}$ alloy film and the $Al/Ni_{80}Fe_{20}$ interface conductance $G_1$ as the only two free parameters, as shown in Fig. 5 (a), and derived $\Lambda_{cross}=22\pm2$ W m$^{-1}$ K$^{-1}$, and $G_1=500$ MW m$^{-2}$ K$^{-1}$. We are able to derive both $\Lambda_{cross}$ of the film and the interface conductance $G_1$ because they affect $R_{dual}$ as a function of delay time in different manners: $\Lambda_{cross}$ mainly affects the amplitude while $G_1$ mainly affects the gradient of $R_{dual}$. We note that our measurements have little sensitive to $G_1$, see Fig. 2(b). Thus, the fitted value of $G_1=500$ MW m$^{-2}$ K$^{-1}$ has a high uncertainty and could be substantially higher than the real value, considering that the Al film was not deposited in situ and thus $G_1$ is limited by the native oxide layer on the NiFe film. This however does not affect the uncertainty of the derived $\Lambda_{cross}$; we varied $G_1$ from 200-600 MW m$^{-2}$ K$^{-1}$, and the derived $\Lambda_{cross}$ is within the 10% uncertainty.

Using this value of thermal conductivity, calculations of the thermal model do not agree with the TDTR measurements at individual frequencies of $f_h=17.4$ MHz and $f_0=4$ MHz, see Fig. 5 (b). This is due to errors in the thickness of Al film which $R_{dual}$ is not sensitive to. Without the dual-frequency approach, fitting of conventional TDTR measurements on the NiFe sample at $f_h=17.4$ MHz yields $\Lambda_{cross}=28\pm6$ W m$^{-1}$ K$^{-1}$ and $G_1=250$ MW m$^{-2}$ K$^{-1}$.

To verify the dual-frequency TDTR measurement, we independently measure the electrical resistivity $\rho=32.3\pm4.1$ μΩ-cm of the NiFe film by a four-point probe. We derive the thermal conductivity $\Lambda$ from electrical resistivity $\rho$ using Wiedemann-Franz law; $\Lambda=LT/\rho$, where $L$ is the Lorenz number for the metal film. Here instead of using the Sommerfeld value for $L$, we estimate the Lorenz number specifically for $Ni_{80}Fe_{20}$ alloy as $(2.38\pm0.2)\times10^{-8}$ Ω W K$^{-2}$ from its bulk values of the thermal conductivity[20] and the electrical resistivity.[21] We note that ~2-3 W m$^{-1}$ K$^{-1}$ of heat could be carried by phonons in metallic alloys/amorphous.[22] This phononic thermal conductivity is included in our estimation of the thermal conductivity of $Ni_{80}Fe_{20}$ alloy using the experimental Lorenz number, as we do not expect the phononic thermal conductivity in NiFe thin films to be significantly reduced from the bulk. (Phonons are already strongly scattered by the high concentration of electrons). We thus estimate the thermal conductivity of our $Ni_{80}Fe_{20}$ film to be 21.7±3.6 W m$^{-1}$ K$^{-1}$. Very good agreement has been achieved between the four-point measurement and our dual-frequency TDTR approach.

(b) Cu film

We also test our dual-frequency approach on a 3-μm-thick Cu film. Based on the general guidelines outlined in Section II (B), we choose $f_h$=9.8 MHz and $f_0$=1.82 MHz for the dual-frequency TDTR measurement of our Cu film. We fit $R_{dual}$ with $\Lambda_{cross}$ of the Cu film and the Al/Cu interface conductance $G_1$ as the only two free parameters, as shown in Fig. 5 (c), and derived $\Lambda_{cross}$=215±26 W m$^{-1}$ K$^{-1}$, and $G_1$=150 MW m$^{-2}$ K$^{-1}$. We note that our measurements have little sensitive to $G_2$. We varied $G_2$ from 50-500 MW m$^{-2}$ K$^{-1}$, and the derived $\Lambda_{cross}$ is within the 5% uncertainty.

Using this value of thermal conductivity, calculations of the thermal model do not agree with the TDTR measurements at individual frequencies of $f_h$=9.8 MHz and $f_0$=1.82 MHz, see Fig. 5 (d). Without the dual-frequency approach, fitting of conventional TDTR measurements on the Cu film sample at $f_h$=9.8 MHz yields $\Lambda_{cross}$=176±37 W m$^{-1}$ K$^{-1}$ and $G_1$=120 MW m$^{-2}$ K$^{-1}$.

To verify the dual-frequency result, we independently measure the electrical resistivity $\rho$ = 2.98±0.33 μΩ-cm of the Cu film by a four-point probe. We also use the Wiedemann-Franz law to estimate the thermal conductivity $\Lambda = LT/\rho$; here the Lorenz number $L$ is taken as 2.20×10$^{-8}$ Ω W K$^{-2}$ for Cu.[23] We thus estimate the thermal conductivity of our Cu film to be 216±32 W m$^{-1}$ K$^{-1}$. The thermal conductivity of the Cu film derived from the dual-frequency TDTR also compares well with the thermal conductivity estimated from the electrical resistivity of the film.

Table 1 summarizes the measurements of thermal conductivity of both the $Ni_{80}Fe_{20}$ alloy film and the Cu film by these different approaches. Good agreements between the thermal conductivity derived from the dual-frequency approach and from the Wiedemann-Franz law validate our dual-frequency TDTR approach.

## IV. SUMMARY

A dual-frequency TDTR approach has been proposed to improve the accuracy of measurements up to ≈3 times and extend the film thickness limit of cross-plane thermal conductivity of high-thermal-conductivity thin films on low-thermal-conductivity substrates up to ≈1.8 times. In the dual-frequency approach, we perform two TDTR measurements, one at a high modulation frequency $f_h$, chosen such that $d_p=0.5h_f$, and another at a low modulation frequency $f_0$, chosen such that $d_p=1.5h_f$. By analyzing the ratio of measurements at $f_h$ to that at $f_0$, we successfully reduce the sensitivity of the measurements to the thickness of Al film, the largest source of uncertainty in TDTR measurements, and thus improve the accuracy by ≈3 times. We show that by using the dual-frequency approach, the minimum film thickness measurable by TDTR is extended from $h_f=1.5d_p$ to $h_f=0.85d_p$. We show that our dual-frequency approach is useful when the thermal conductance of the transducer/film interface is sufficiently high ($G_1>\Lambda_{cross}/h_f$) and the thermal conductivity of the substrate is sufficiently low ($\Lambda_{cross}/\Lambda_s>10$). We verify the dual-frequency approach by measuring the thermal conductivity of a 400-nm-thick $Ni_{80}Fe_{20}$ alloy film and a 3-μm-thick Cu film on thermal $SiO_2$. The measurements by our dual-frequency approach agree favorably with independent four-point probe measurements, with the uncertainty reduced from 21% to ~10%.

## ACKNOWLEDGMENTS

We are grateful to Prof T. Osipowicz and Dr. M. Ren (National University of Singapore) for conducting RBS and PIXE measurements on our $Ni_{80}Fe_{20}$ alloy samples. This material is based upon work supported by the NUS Start-Up Grant.

Table 1 Summary of measured thermal conductivity of $Ni_{80}Fe_{20}$ alloy film and Cu film by three different approaches

|  | Four-point | TDTR@17.4 MHz | Dual-frequency TDTR |
|---|---|---|---|
| $\Lambda_{NiFe}$ (W m$^{-1}$ K$^{-1}$) | 21.7±3.6 | 28±6 | 22±2 |
| $\Lambda_{Cu}$ (W m$^{-1}$ K$^{-1}$) | 216±32 | 176±37 | 215±26 |

## Figures:

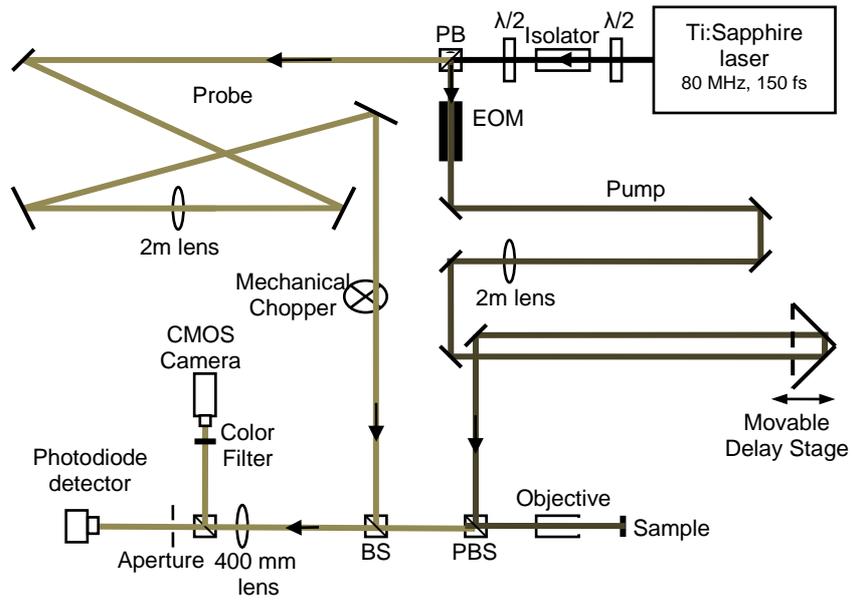

Figure 1 A schematic diagram of our TDTR setup. EOM represents electro-optical modulator; PBS represents polarizing beam splitter; BS represents beam splitter; and $\lambda/2$ represents half wave plate.

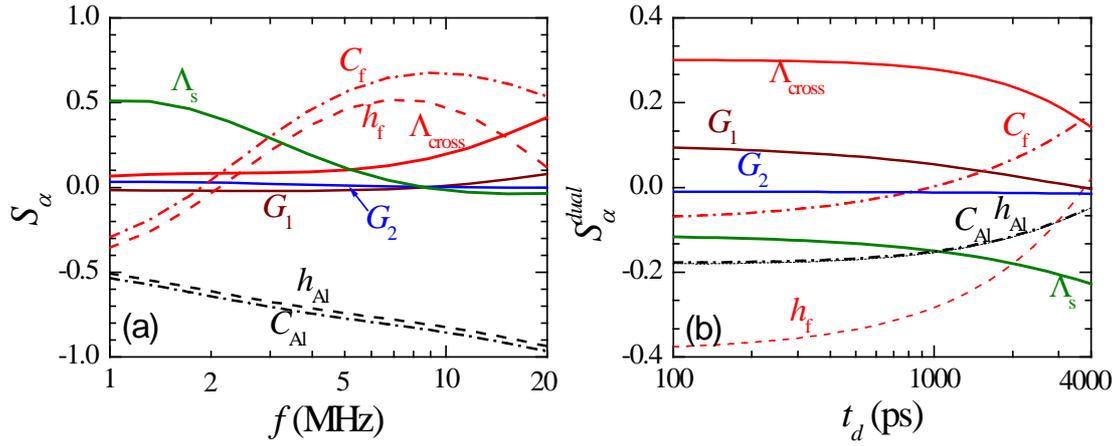

Figure 2 (a) Sensitivity of TDTR signals to different parameters in the thermal model, as a function of modulation frequency. The sample is a hypothetical, multilayered structure of 100 nm Al / 500 nm thin film / 100 nm SiO$_2$ / Si substrate, with the film thermal diffusivity $D_f=10^{-5}$ m$^2$ s$^{-1}$. The parameters include heat capacity $C_{Al}$ and thickness $h_{Al}$ of the Al layer, heat capacity $C_f$, thickness $h_f$ and cross-plane thermal conductivity $\Lambda_{cross}$ of the hypothetical film, thermal conductivity $\Lambda_s$ of underlying SiO$_2$ film, thermal conductance $G_1$ of Al/film interface and $G_2$ of film/SiO$_2$ interface. We assume $C_f=2$ J cm$^{-3}$ K$^{-1}$, $G_1=200$ MW m$^{-2}$ K$^{-1}$, $G_2=200$ MW m$^{-2}$ K$^{-1}$ and laser spot $1/e^2$ radii of 28 μm, and fix the delay time at 100 ps. (b) Sensitivity of the ratio of TDTR signals at 20 MHz and 6 MHz of the hypothetical sample, plotted as a function of delay time.

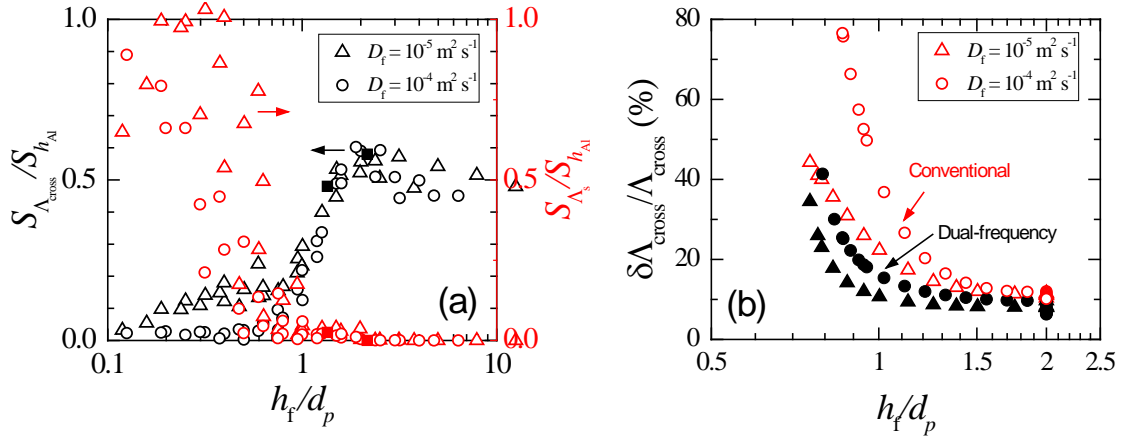

Figure 3 (a) The sensitivity of TDTR signals to $\Lambda_{cross}$ of the sample film and the sensitivity to $\Lambda_s$ of the underlying insulating layer respectively, normalized by the sensitivity to $h_{Al}$, of the hypothetical sample described in Fig. 2, plotted as a function of film thickness $h_f$ normalized by the thermal penetration depth $d_p$. In this plot, we change the thickness (0.3-10 μm) and the thermal diffusivity (open triangles for 10$^{-5}$ m$^2$ s$^{-1}$ and open circles for 10$^{-4}$ m$^2$ s$^{-1}$) of the hypothetical film, and assume that the measurements are performed at a modulation frequency ranging from 0.1 MHz to 20 MHz. The heat capacity of the film $C_f$ is kept constant as 2 J cm$^{-3}$ K$^{-1}$. The solid square symbols are for the Ni$_{80}$Fe$_{20}$ alloy sample and the Cu film sample that we measured in this study. (b) The estimated uncertainties of $\Lambda_{cross}$ of the hypothetical thin film derived by the dual-frequency TDTR approach (solid) compared to that derived by the conventional TDTR approach (open) for thermal diffusivity of the film of 10$^{-5}$ m$^2$ s$^{-1}$ (triangles) and 10$^{-4}$ m$^2$ s$^{-1}$ (circles). The conventional TDTR is assumed to be performed at a modulation frequency of $f_h$ such that $d_p = 0.5 h_f$, or 20 MHz, whichever is smaller, while the dual-frequency TDTR is assumed to be performed according to the general guideline presented in the text. The calculations are plotted as a function of film thickness, normalized by the thermal penetration depth at a modulation frequency of $f_h$. The uncertainties of $\Lambda_{cross}$ are estimated using Eq. (2). Among the input parameters, the uncertainties of heat capacities are estimated as 3%, thicknesses (except substrate) estimated as 5%, thermal conductivity (except the target film) estimated as 10%, the front and back interface conductance estimated as 20%, and the laser spot size estimated as 5%. The uncertainty of the phase is estimated assuming signal-to-noise ratio of TDTR signal as 100.

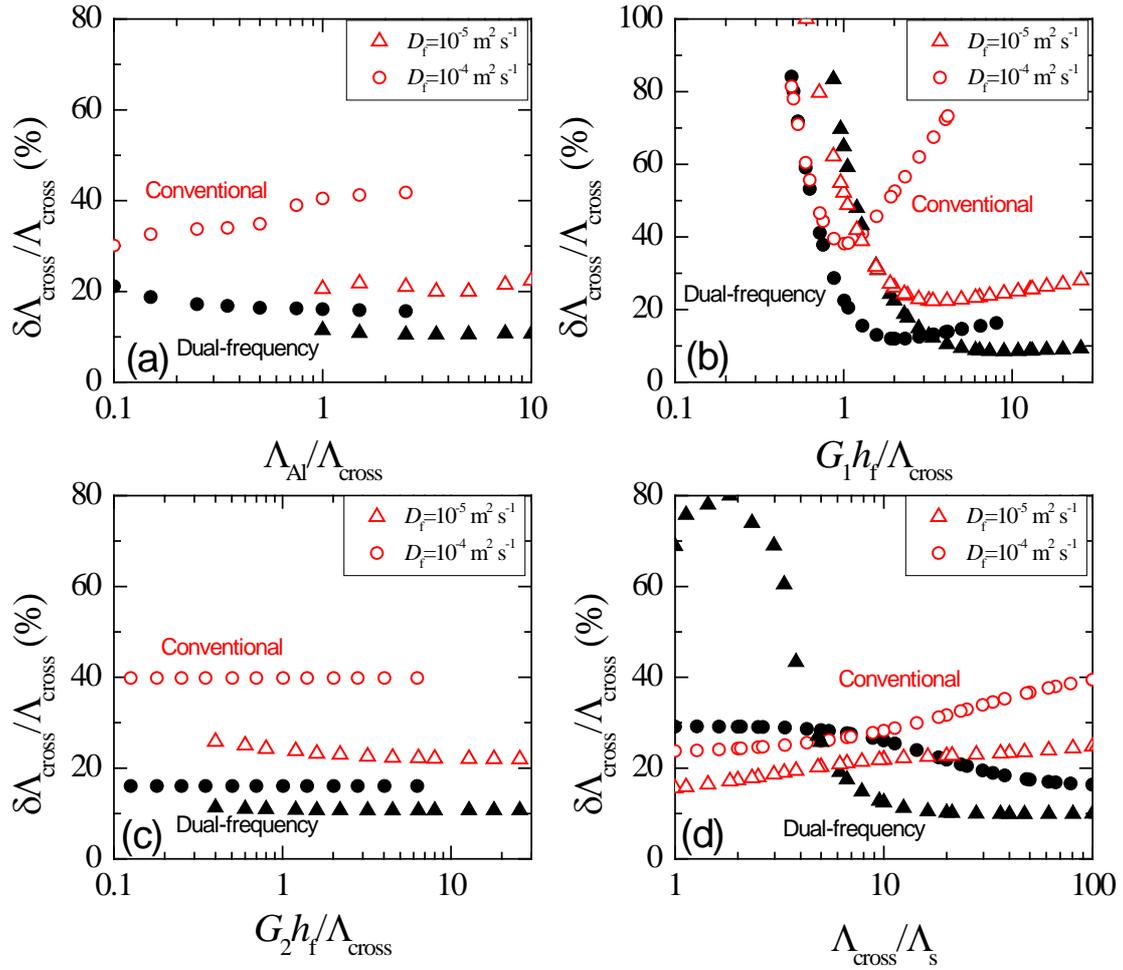

Figure 4 The estimated uncertainties of $\Lambda_{cross}$ of the hypothetical thin film as described in Fig. 2, derived by the dual-frequency TDTR approach (solid) compared to that derived by the conventional TDTR approach (open), for thermal diffusivity of the film of $10^{-5}$ m$^2$ s$^{-1}$ (triangles) and $10^{-4}$ m$^2$ s$^{-1}$ (circles), respectively. The film thickness $h_f$ was chosen such that $d_p = h_f$ at a modulation frequency $f_h$ of 20 MHz; $h_f$=400 nm for the film with thermal diffusivity of $10^{-5}$ m$^2$ s$^{-1}$ and $h_f$=1260 nm for the film with thermal diffusivity of $10^{-4}$ m$^2$ s$^{-1}$. The conventional TDTR is assumed to be performed at 20 MHz, while the dual-frequency TDTR is assumed to be performed according to the general guideline presented in the text. The calculations are performed using the properties described in Fig. 2, except the following: (a) we systematically vary the thermal conductivity of the Al film, $\Lambda_{Al}$, and plot the uncertainty as a function of $\Lambda_{Al}/\Lambda_{cross}$, (b) we systematically vary the thermal conductance of Al/film interface $G_1$, and plot the uncertainty as a function of dimensionless $G_1 h_f/\Lambda_{cross}$, (c) we systematically vary the thermal conductance of film/substrate interface $G_2$, and plot the uncertainty as a function of dimensionless

$G_2 h_f/\Lambda_{cross}$, and (d) we systematically vary the thermal conductivity of the substrate $\Lambda_s$, and plot the uncertainty as a function of $\Lambda_{cross}/\Lambda_s$. The uncertainties of $\Lambda_{cross}$ are estimated in the same manner as described in Fig. 3(b).

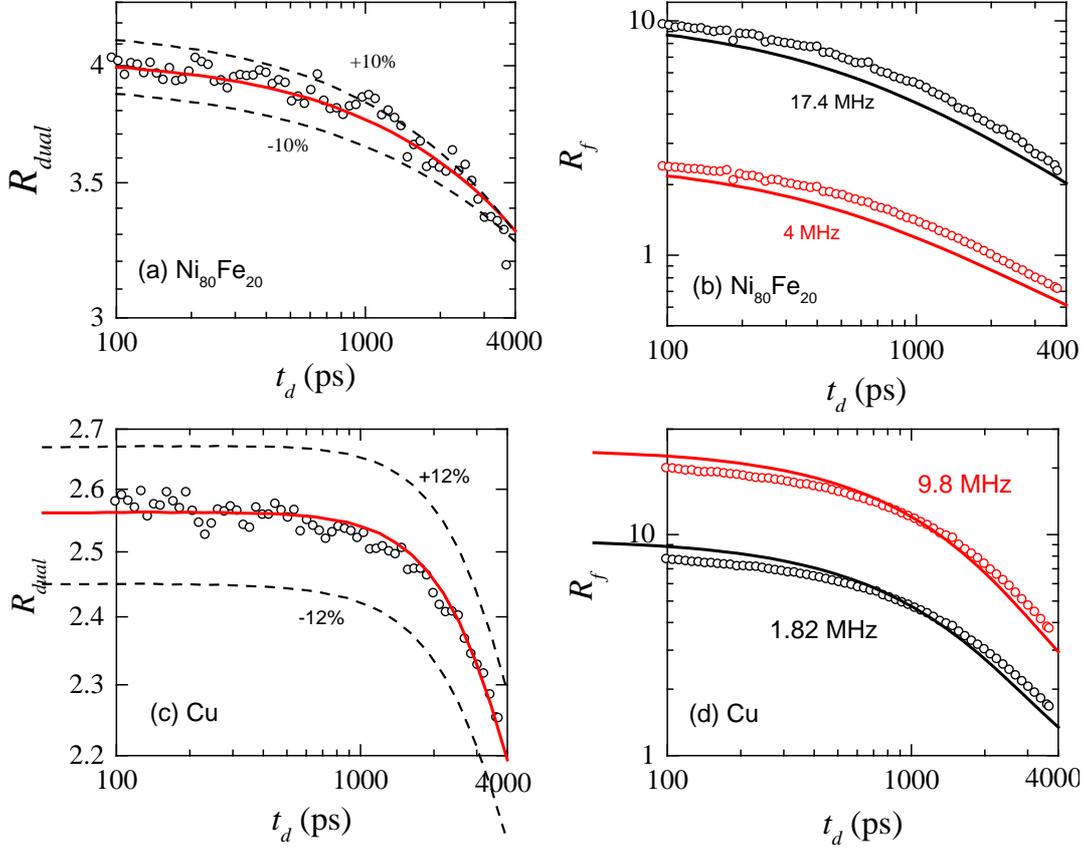

Figure 5 (a) Fitting of $R_{dual}$ of the $Ni_{80}Fe_{20}$ alloy sample, measured with laser spot $1/e^2$ radii of 28 µm at frequencies $f_h$ as 17.4 MHz and $f_0$ as 4 MHz, yielding $\Lambda_{cross}$ of the alloy film as 22 W m$^{-1}$ K$^{-1}$(solid line), with 10% bounds on the fitted value (dashed lines). (b) The thermal model with fitted values of thermal conductivity from (a) could not fit either of the TDTR signals $R_f$ at the two frequencies. (c) Fitting of $R_{dual}$ of the Cu film sample, measured with laser spot $1/e^2$ radii of 10 µm at frequencies $f_h$ as 9.8 MHz and $f_0$ as 1.82 MHz, yielding $\Lambda_{cross}$ of the Cu film as 215 W m$^{-1}$ K$^{-1}$(solid line), with 12% bounds on the fitted value (dashed lines). (d) The thermal model with fitted values of thermal conductivity from (c) could not fit either of the TDTR signals $R_f$ at the two frequencies.